\documentclass{WileyMSP-template}

\usepackage[backend=biber]{biblatex}
\DeclareNameAlias{sortname}{family-given} % no effect for sorting=none

\usepackage{geometry}

\usepackage[dvipsnames]{xcolor}

\usepackage{acro} % better than acronym

\usepackage{amsmath}
\usepackage{amssymb}
\usepackage{esdiff}
\usepackage{leftindex}
\usepackage{bm}

\usepackage{multirow}

\usepackage{subeqnarray}
\usepackage{caption}
\usepackage[labelfont=footnotesize,textfont=footnotesize]{subcaption}

\usepackage{tikz}
\usetikzlibrary{arrows,shapes,fadings,backgrounds,decorations.shapes,patterns}

\usepackage{verbatim}

\usepackage{ragged2e}

\usepackage{longtable}
\usepackage{booktabs}

\usepackage{miller}
\usepackage{siunitx}

\usepackage[toc]{appendix}

\usepackage{graphicx}

\usepackage{siunitx}

\usepackage{textgreek}
\usepackage{upgreek}

\usepackage{xargs}

\usepackage[pdfencoding=auto,
bookmarksnumbered=true,%                                % generate bookmarks with numbers
colorlinks=true,%
linkcolor=blue,
citecolor=blue,
urlcolor=blue,
pdfpagelabels=false,
plainpages = false,
]{hyperref}%

\usepackage[capitalise]{cleveref}                       % for clever referencing

\newcommand{\vctr}[1]{\ensuremath{\mathbf{#1}}}

\begin{document}

\pagestyle{fancy}

\title{Incorporation of physics-based strengthening coefficients into phenomenological crystal plasticity models}

\maketitle

% Author: Please give full first and last names for authors and include * after the name of all corresponding authors

\author{Nikhil Prabhu}
\author{Martin Diehl*}

% Affiliations: Please provide adacemic titles (Prof. or Dr.) for all authors where applicable, and include an institutional email address for all corresponding authors
\begin{affiliations}
N. Prabhu, M. Diehl\\
Department of Materials Engineering, KU Leuven, Kasteelpark Arenberg 44, 3001 Leuven, Belgium\\
Department of Computer Science, KU Leuven, Celestijnenlaan 200A, 3001 Leuven, Belgium\\
E-mail: martin.diehl@kuleuven.be
\end{affiliations}

% Keywords: Please provide a minimum of three and a maximum of seven keywords, separated by commas


% Abstract should be written in the present tense and impersonal style (i.e., avoid we), and be at most 200 words long
\begin{abstract}
The efforts associated with parametrization of continuum-based models for crystal plasticity are a significant obstacle for the routine use of these models in materials science and engineering.
While phenomenological constitutive descriptions are attractive due to their small number of adjustable parameters, the lack of physical meaning of their parameters counteracts this advantage to some extent.
This study shows that interaction/strengthening coefficients determined with the help of discrete dislocation dynamics simulations for use in physics-based formulations can also be used to improve the predictive quality of phenomenological models.
Since the values of these parameters have been determined for most technologically relevant materials, the findings enable to improve the parametrization of phenomenological crystal plasticity models at no costs.
\end{abstract}

\tableofcontents

\section{Introduction}
Continuum crystal plasticity models are typically classified as either \emph{phenomenological} or \emph{physics-based} \cite{RotersEtAl2010,Diehl2024,PatilEtAl2021}.
While former models rely solely on fitting the parameters of an ad hoc chosen functional form, physics-based models make---at least partly---use of explainable parameters such as dislocation density, activation energy, or Burgers vector.
Although physics-based models in general do not lead to better predictions \cite{PrabhuEtAl2024}, the use of explainable parameters and mechanism-based equations gives them two distinct advantages:
1) they have predictive capabilities outside of the data range used to determine their parameters 2) they can be formulated and parameterized with the help of simulations methods for smaller scales such as density functional theory (DFT) \cite{WongEtAl2016} and discrete dislocation dynamics (DDD) \cite{StrickerEtAl2018,AkhondzadehEtAl2020}.
DDD simulations are particularly useful for the the determination of parameters that describe dislocation--dislocation interactions \cite{
MadecEtAl2003,% early measurements
KubinEtAl2008,%
QueyreauEtAl2009,%
BertinEtAl2014,%
MadecEtAl2017%
} which are the reason for strain hardening.
%Despite these advantages of physics-based models, phenomenological models are widely used.

The primary state variable of physics-based constitutive models is the dislocation density $\rho$, hence they are also called dislocation density-based models \cite{PatilEtAl2021}.
A plethora of formulations has been proposed to describe the dislocation population and predict its evolution during deformation.
While simple models rely on the evolution of a generic dislocation density per slip system \cite{MeckingEtAl1981}, sophisticated approaches are based on multiple dislocation densities and consider their characteristics for mechanism-based evolution rules \cite{RotersEtAl2000,ReuberEtAl2014}.
Strain hardening in most of these models is derived from the so-called Taylor relation, which implies that the critical resolved shear stress $\tau_\mathrm{crit}$ (CRSS) is proportional to the square root of the dislocation density.
Extensions of the Taylor relation that consider dislocation densities per slip system proposed by
\textcite{LavrentevEtAl1975a} and \textcite{FranciosiEtAl1980} predict system-dependent hardening and are typically used in crystal plasticity models.
The latter formulation, which has demonstrated its ability to reproduce results from DDD simulations \cite{QueyreauEtAl2009,BertinEtAl2014}, reads for $N$ slip systems
\begin{equation}
	\tau_\mathrm{crit}^i = \mu b \sqrt{\left(\sum_j^N a^{ij} \rho^j \right)},
    \label{dislocation}
\end{equation}
with shear modulus $\mu$ and magnitude of the Burgers vector $b$.
The entries $a^{ij}$ of the interaction matrix $\vctr{a}$ are a measure of the average interaction strength that dislocations on system $j$ exert on dislocations from system $i$.
The values of most of these \emph{interaction coefficients} can be obtained directly from DDD simulations \cite{MadecEtAl2003} in dependence of material \cite{MadecEtAl2017} and dislocation density \cite{DevincreEtAl2006,KubinEtAl2008}.

In phenomenological models, the resistance against plastic flow $\tau_\mathrm{crit}$, i.e. the CRSS per slip system, is directly used as a state variable.
Various hardening rules have been proposed to describe the evolution of the CRSS, e.g. the extension of the Voce law \cite{Voce1948,Voce1955} to multiple slip systems \cite{TomeEtAl1984,BartonEtAl2001}.
Another commonly chosen form is a power law \cite{PeirceEtAl1982}, % eq 3.35
\begin{equation}
	\dot{\tau}_\mathrm{crit}^i = \hslash \left| 1 - \frac{\tau^i_\mathrm{crit}}{\tau_\mathrm{crit}^\infty}\right|^a \sum_j^N h^{ij} \left| \dot{\gamma}^j \right|,
    \label{pheno}
\end{equation}
where $\hslash$ and $a$ are fitting parameters, $\tau_\mathrm{crit}^\infty$ is the saturation value of $\tau_\mathrm{crit}$, $\dot{\gamma}$ is the shear rate, and $\vctr{h}$ is the \emph{hardening moduli} matrix \cite{Hutchinson1970,PeirceEtAl1983}.
Despite the clear analogy between $\vctr{a}$ and $\vctr{h}$, typically $h^{ij}$ are chosen to be either $1.0$ or $1.4$ \cite{FrydrychEtAl2022,FischerEtAl2025}.
This choice dates back---to the best knowledge of the authors---to a limited set of experimental observations reported by \textcite{Kocks1970} which motivated \textcite{PeirceEtAl1982} to use values of \num{1.0} for all coplanar interactions and \num{1.4} otherwise.
A simplification of this parametrization is frequently used where \num{1.0} is used only for self interactions and not for other coplanar interactions which also have a value of \num{1.4} \cite{KhadykoEtAl2014}.
While the origins of this simplification are unclear, it should be noted that there is strong evidence that the coplanar interaction is weak, in particular in comparison to the strong collinear interaction \cite{MadecEtAl2003}.

In this study, it is investigated whether the parametrization of phenomenological models is improved when values obtained for the interaction in physics-based models, i.e. for $a^{ij}$, are used for $h^{ij}$ instead of the ubiquitous \num{1.0}/\num{1.4} combination.
To this end, reference results are computed with a dislocation density-based model and compared to results from a phenomenological model with different choices for entries of the hardening moduli matrix.

\section{Simulation Setup}
All simulations are performed using DAMASK 3.0.1 \cite{RotersEtAl2019,DiehlEtAl2020} with a Fast Fourier Transform-based spectral solver \cite{ShanthrajEtAl2015,ShanthrajEtAl2019,Schneider2021}.
To avoid complications arising when more than one family of slip systems exists, which might have  different hardening behavior, a face-centered cubic (fcc) lattice is assumed.
In the case of fcc crystals that deform exclusively on the 12 octahedral \hkl{111}\hkl<110> slip systems, there exists 7 different interaction mechanisms.
Following the notation of \textcite{MadecEtAl2017}, these are labeled as \emph{self} ($h_0$), \emph{coplanar} ($h_\mathrm{copla}$), and \emph{collinear} ($h_\mathrm{coli}$) interactions; \emph{Hirth lock} ($h_1$), glissile junctions $G_{\qty{0}{\degree}}$ and $G_{\qty{60}{\degree}}$ ($h_2$, $h_2^*$), and \emph{Lomer lock} ($h_3$).

\subsection{Microstructures and Textures}
Two different microstructures are synthetically generated, one with globular grains and a random texture and one with flattened grains (aspect ratio 1:1:3) and an idealized fcc rolling texture.
Both microstructures are discretized by 48$\times$48$\times$48 voxels, contain 1000 grains, and are periodically repeated at the boundaries.
The rolling texture was created using texture components with Gaussian scatter following the ideas of \textcite{Helming1998}.
It consists of \qty{30}{\percent} S orientation ($\hkl{123}\hkl<634>$), \qty{20}{\percent} Cu orientation (\hkl{112}\hkl<111>),  \qty{15}{\percent} Brass orientation (\hkl{011}\hkl<211>), \qty{10}{\percent} \textalpha-fiber (\hkl<110> || ND), and a random background \cite{Ray1995,KestensEtAl2016}.
A scatter with a full width at half maximum of \qty{50}{\degree} and \qty{20}{\degree} was used for the texture components and for the fiber, respectively.

\subsection{Load Cases}
The considered load cases are uniaxial tension and simple shear, both to a strain of approximately\footnote{The load is defined in terms of the deformation gradient $\mathbf{F}$ and a final value of $1.15$ and $0.15$ was prescribed for respective direction of the tensile ($F_{11}$) and shear ($F_{21}$) load, respectively.} \qty{15}{\percent} at a strain rate of \qty{1e-3}{\per \second}; the two (uniaxial tension) or three (simple shear) undefined normal components of stress are set to \qty{0.0}{\mega \pascal} to enable volume changes due to contraction in the elastic regime.
These values are volume averages and periodic boundary conditions, which are inherent to the spectral method, are applied.

\subsection{Reference Results}
A dislocation density-based model is used to obtain a ground truth benchmark.
In this model the interactions are described by \cref{dislocation}, i.e. the values of the interaction matrix are independent of the dislocation density.
An existing parametrization for Inconel 625 \cite{PrabhuEtAl2024}, a nickel-based superalloy, which employs interaction coefficients for Nickel from \cite{MadecEtAl2017}, is used.

\subsection{Parameter Study}

\begin{table}
    \centering
    \caption{Independent values of the hardening moduli matrix $\vctr{h}$ of the phenomenological model used in the parameter study.
    The strengthening and interaction coefficients are taken from \cite{MadecEtAl2017} and scaled such that $h_0=1.0$.}
    \label{h values}
    \begin{tabular}{rl l l l l l l}
        &$h_0$&$h_\mathrm{copla}$&$h_\mathrm{coli}$&$h_1$&$h_2$&$h_2^*$&$h_3$ \tabularnewline
        \toprule
        traditional & 1.0  & 1.0  & 1.4  & 1.4  & 1.4  & 1.4  & 1.4 \tabularnewline
        simplified & 1.0  & 1.4  & 1.4  & 1.4  & 1.4  & 1.4  & 1.4 \tabularnewline
        strengthening coeff. & 1.00 & 1.00 & 2.35 & 0.60 & 0.91 & 0.86 & 1.17 \tabularnewline
        interaction coeff. & 1.00 & 1.00 & 5.51 & 0.36 & 0.84 & 0.74 & 1.38 \tabularnewline
        \bottomrule
    \end{tabular}
\end{table}

A phenomenological power law formulation is parameterized with the following four choices for the hardening moduli matrix in \cref{pheno}:

\begin{itemize}
	\item \textbf{Simplified}: All moduli have a value of \num{1.4} except for the self interaction which has a value of 1.0.
	\item \textbf{Traditional}: Self and coplanar interactions have a value of \num{1.0}; the value of all remaining interactions is set to \num{1.4}.
    \item \textbf{Strengthening coefficients} ($\alpha$): Strengthening coefficients are directly obtained from DDD simulations \cite{MadecEtAl2017,KubinEtAl2008} and are scaled such that $h_0=1.0$.
    \item \textbf{Interaction coefficients} ($a$): These values are used in dislocation density-based constitutive laws (and constitute the interaction matrix $a^{ij}$ that appears in \cref{dislocation}) and are the square of the strengthening coefficients \cite{MadecEtAl2017}.
\end{itemize}
The resulting independent values for the four cases are shown in \cref{h values}.

The parameters of this model, more precisely $\hslash$, $\tau_\mathrm{crit}^\infty$, and $a$ from \cref{pheno}, are adjusted for all four cases individually such that the average stress--strain behavior of the reference material is reproduced.
To this end a simplified polycrystal model consisting of $10 \times 10 \times 10$ material points with 1000 grains is used, i.e. each ``grain'' discretized by a single voxel.
The parameters are optimized using the Nelder--Mead algorithm available in SciPy \cite{NelderEtAl1965,GaoEtAl2010,VirtanenEtAl2020}.
The loss is the difference in stress in loading direction from all four combinations of texture (random and rolling) and load case (uniaxial tension and simple shear) which are considered with equal weights.

\subsection{Measuring Correlation}
\label{measuring correlation}
The agreement between the reference solution and the different parametrizations of the phenomenological model can be conveniently visualized in correlation plots.
For a perfect agreement, all points are located on a line with a slope of \num{1.0} (\qty{45}{\degree} inclination) that goes through the origin which is called the \emph{1:1 line}.
For an imperfect agreement, two kind of deviations from this ideal situation can be observed:
1) the best-fit line might deviate from this line.
This is called a lack of \emph{accuracy} \cite{Lin1989}.
2) the individual points scatter around the best-fit line.
This is called a lack of \emph{precision} \cite{Lin1989}.

The concordance correlation coefficient $\rho_{\text{c}}$ proposed by \textcite{Lin1989} is used in this study to measure to which degree data pairs fall on the 1:1 line.
It accounts conveniently for both, precision and accuracy, and is defined as:
\begin{equation}
  \rho_{\text{c}} := 1 -\frac{\mathbb{E}[\epsilon^2]}{\mathbb{E}[\epsilon^2] \vert_{\rho=0}}.
  \label{concordance correlation coefficient}
\end{equation}
In \cref{concordance correlation coefficient} the numerator $\mathbb{E}[\epsilon^2]$ is the expected square of the distance from the 1:1 line.
The denominator is the same measure but for the case that the two variables are uncorrelated, i.e. Pearson's correlation coefficient $\rho$ is zero.
After simplification, $\rho_{\text{c}}$ can be computed using standard deviation ($\sigma_{z}$ and $\sigma_{\hat{z}}$), mean ($\mu_{z}$ and $\mu_{\hat{z}}$) and covariance ($\sigma_{z \hat{z}}$) as follows:
\begin{equation}
    \rho_{\text{c}} = \frac{2 \sigma_{z \hat{z}}} {{\sigma_{z}^{2}} +  {\sigma_{\hat{z}}^{2}} + {\left(\mu_{z} - \mu_{\hat{z}}\right)}^2} = \frac{\sigma_{z \hat{z}}}{\sigma_{z} \sigma_{\hat{z}}} \ \frac{2}{\frac{1}{\sigma^{\ast}} + \sigma^{\ast} + u^2} =: \rho C_{\text{b}},
    \label{CCC}
\end{equation}
where $\sigma^{\ast} = \frac{\sigma_{\hat{z}}}{\sigma_{z}}$, $u = \frac{\mu_{z} - \mu_{\hat{z}}}{\sqrt{\sigma_{z} \sigma_{\hat{z}}}}$, and $z$ and $\hat{z}$ denote material point data from physics-based and phenomenological models, respectively.
\textcite{Lin1989} multiplicatively decomposed $\rho_{\text{c}}$ into two contributions as seen in \cref{CCC}: Pearson's correlation coefficient $\rho$ which measures the degree of linear correlation, and bias correction factor $C_{\text{b}}$ which measures how far the best-fit line deviates from the 1:1 line.
Interpretation of value of $\rho_{\text{c}}$ is analogous to that of $\rho$: \num{1} indicates perfect agreement ($z = \hat{z}$), \num{-1} indicates perfect disagreement ($z = -\hat{z}$), and a departure from these terminal values towards zero indicate lowering degree of agreement.
Although $\rho_{\text{c}}$ is commonly used as a single index when comparing multiple models against a sole reference dataset \cite{WadouxEtAl2024}, the decomposition products $\rho$ and $C_{\text{b}}$ are also shown.
In addition, the  modeling efficiency coefficient \cite{JanssenEtAl1995} $\text{MEC}:= 1 - \frac{\text{MSE}}{\sigma_{z}^2}$ where $\text{MSE}= \sum_{i=1}^{N} \left( z - \hat{z} \right)^2$, is calculated.

\section{Results and Discussion}
\begin{figure}
    \centering
    \includegraphics[width=.60\textwidth]{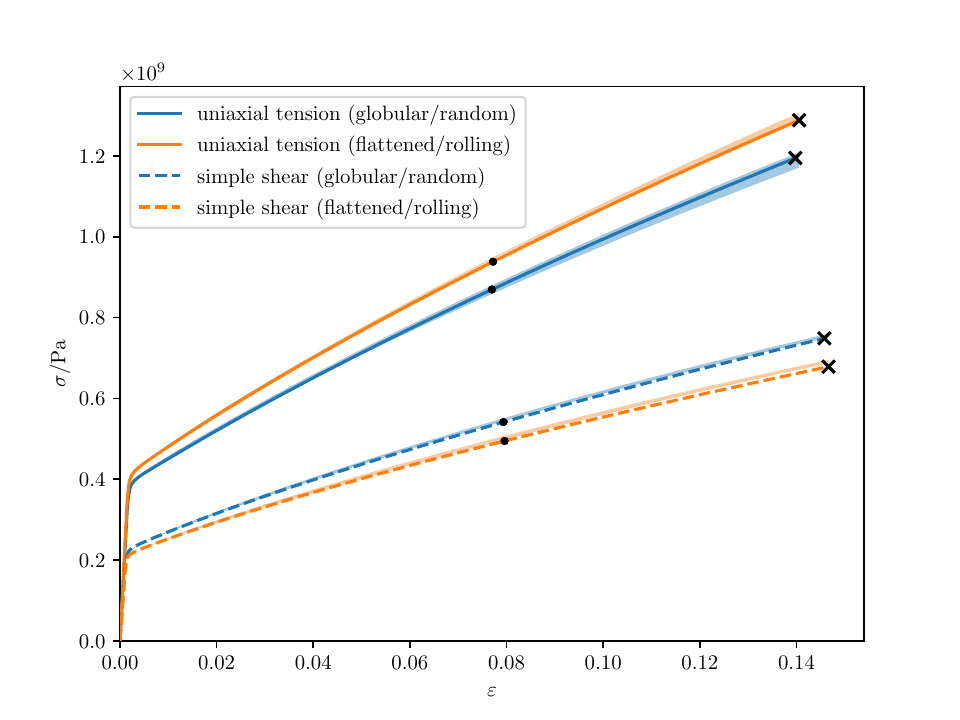}
    \caption{Stress--strain curve in loading direction for the four combinations of microstructure/texture and load. The solid and dashed lines indicate the reference results obtained with the dislocation density-based model and the shaded background represents the range obtained from the four different parametrizations of the phenomenological model. The black dots mark strain levels of approximately \qty{8}{\percent}, the black crosses the final load of approximately \qty{15}{\percent}.}
    \label{stress-strain}
\end{figure}

As a prerequisite for an objective comparison of the different parametrizations, it has to be established that the phenomenological model reproduces the macroscopic behavior of the dislocation density-based model independently of the chosen parametrization.
To this end, the equivalent stress--strain curves for both microstructures loaded in tension and in simple shear are shown in \cref{stress-strain}.
The reference results are shown as a solid line and the minimum and maximum among all four parametrizations is shown as a shaded area.
It can be seen that all parametrizations of the phenomenological constitutive law reproduce the macroscopic behavior of the physics-based formulation.
In this plot, the strain levels ($\varepsilon \approx 0.08$ and $\varepsilon \approx 0.15$) at which the correlation is investigated are also shown.

The correlation between the reference solution and the different parametrizations is determined independently for the hardening behavior and for the deformation behavior.
For the hardening behavior, the CRSS increase per slip system of all material points, $\Delta \tau_\mathrm{crit} := \tau_\mathrm{crit} - \tau_{\mathrm{crit},0}$, is examined.
For the deformation behavior, the accumulated shear $\gamma$ per slip system is examined.
Since it is known only a few slip systems need to be activated to achieve the applied deformation \cite{Taylor1938}, only active slip systems, i.e. systems with $\gamma > \num{1e-3}$ are considered for this analysis.
\Cref{fraction active} gives the fraction of active slip systems for which $\gamma > \num{1e-3}$ and shows that about \qty{30}{\percent}, i.e. 4 out of the 12 slip systems, are active for the considered cases.

\begin{table}
    \caption{Fraction of slip systems that are active, i.e. have accumulated slip $\gamma > \num{1e-3}$ for the different parametrizations.
    The values given in normal font are the average of all four combinations of microstructure/texture and load which are complemented by their minimum in subscript and their maximum in superscript.}
    \label{fraction active}
    \centering
    \begin{tabular}{r c c }
        & $\varepsilon \approx 0.08$ & $\varepsilon \approx 0.15$ \tabularnewline
        \toprule
        simplified           & $0.28\,_{0.26}^{0.29}$ & $0.30\,_{0.29}^{0.31}$ \tabularnewline[.6ex]
        traditional          & $0.28\,_{0.26}^{0.29}$ & $0.30\,_{0.29}^{0.31}$ \tabularnewline[.6ex]
        strengthening coeff. & $0.30\,_{0.29}^{0.31}$ & $0.33\,_{0.32}^{0.33}$ \tabularnewline[.6ex]
        interaction coeff.   & $0.31\,_{0.30}^{0.31}$ & $0.33\,_{0.33}^{0.34}$ \tabularnewline
        \bottomrule
    \end{tabular}
\end{table}

\subsection{Hardening Behavior}
\begin{figure}
	\includegraphics[width=\linewidth]{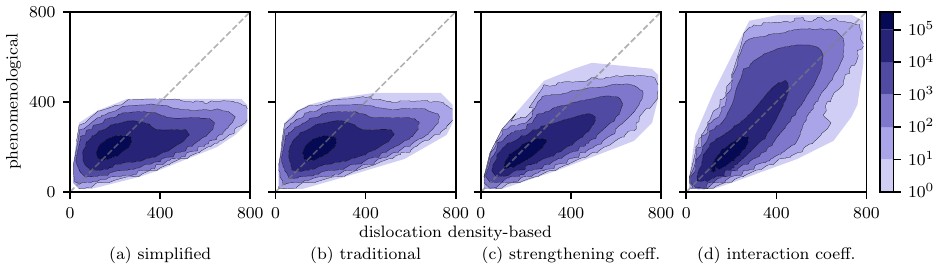}
	\caption{Density plot showing the correlation between the increase in CRSS ($\Delta \tau_\mathrm{crit}$ in \unit{\mega \pascal}) obtained with the different parametrizations of the phenomenological model and the dislocation density-based model at an applied strain of approximately \qty{8}{\percent} in loading direction.}
	\label{DeltaTau 110}
\end{figure}
\begin{figure}
	\includegraphics[width=\linewidth]{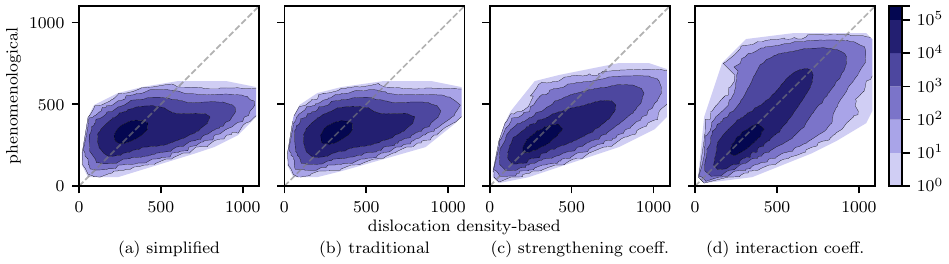}
	\caption{Density plot showing the correlation between the increase in CRSS ($\Delta \tau_\mathrm{crit}$ in \unit{\mega \pascal}) obtained with the different parametrizations of the phenomenological model and the dislocation density-based model at an applied strain of approximately \qty{15}{\percent} in loading direction.}
	\label{DeltaTau 180}
\end{figure}

Correlation plots of the CRSS increase $\Delta \tau_\mathrm{crit}$ per slip system of all material points are shown at applied strains of approximately \qty{8}{\percent} and \qty{15}{\percent} (c.f. \cref{stress-strain}) in \cref{DeltaTau 110,DeltaTau 180}, respectively.
The data is plotted individually for the four considered parametrizations as density contour plots with the color code representing point density in terms of number of data points on a logarithmic scale.
The figures reveal that the traditional and simplified parametrizations perform worse than the physics-informed parametrization and the use of interaction parameters leads to the closest agreement.
Detailed inspection of the raw data \cite{prabhu_2025_14799861} reveals that $\Delta \tau_\mathrm{crit}$ of the primary, i.e. the most active, slip system is in a significantly better agreement than the remaining slip systems for the traditional and simplified parameter sets.
Hence, besides a clear deviation of contours from 1:1 line, a hint of bimodal distribution can be seen in \cref{DeltaTau 110}(a) and (b) and \cref{DeltaTau 180}(a) and (b).
Such behavior is not seen when using the physics-informed parametrizations.
There is also a clear difference between the results from using strengthening coefficients to that from using interaction coefficients:
While the former have a distribution that is located below the 1:1 line, indicating less hardening of the phenomenological model than the reference result, using the interaction coefficients leads to an almost symmetric distribution of high-density contour regions about the line of equality.

\begin{table}
    \caption{Concordance correlation coefficient $\rho_{\text{c}}$ and its decomposed products: Pearson's correlation coefficient $\rho$ and bias correction factor $C_\text{b}$; and modeling efficiency coefficient MEC measuring the degree of agreement between the reference results and the different parametrizations for the increase in CRSS $\Delta \tau$.
    The values given in normal font are the average of all four combinations of microstructure/texture and load which are complemented by their minimum in subscript and their maximum in superscript.}
    \label{fit tau}
	\centering
	\begin{tabular}{r c c c c | c c c c}
                & \multicolumn{4}{c}{$\varepsilon \approx 0.08$} & \multicolumn{4}{c}{$\varepsilon \approx 0.15$}\tabularnewline
                \cline{2-9}
                & $\rho$ & $C_b$ & $\rho_{\text{c}}$ & MEC & $\rho$ & $C_\text{b}$ & $\rho_{\text{c}}$ & MEC \tabularnewline
    \toprule
    simplified           & $0.46\,_{0.44}^{0.51}$ & $0.68\,_{0.66}^{0.70}$ & $0.31\,_{0.29}^{0.34}$ & $0.15\,_{0.12}^{0.17}$ & $0.43\,_{0.40}^{0.46}$ & $0.64\,_{0.63}^{0.66}$ & $0.27\,_{0.26}^{0.29}$ & $0.13\,_{0.11}^{0.15}$ \tabularnewline[.6ex]
    traditional          & $0.45\,_{0.42}^{0.49}$ & $0.69\,_{0.68}^{0.71}$ & $0.31\,_{0.29}^{0.33}$ & $0.13\,_{0.10}^{0.15}$ & $0.42\,_{0.38}^{0.44}$ & $0.65\,_{0.65}^{0.67}$ & $0.27\,_{0.25}^{0.28}$ & $0.11\,_{0.08}^{0.14}$ \tabularnewline[.6ex]
    strengthening coeff. & $0.86\,_{0.86}^{0.88}$ & $0.77\,_{0.74}^{0.80}$ & $0.67\,_{0.65}^{0.68}$ & $0.51\,_{0.46}^{0.54}$ & $0.82\,_{0.81}^{0.84}$ & $0.74\,_{0.71}^{0.76}$ & $0.61\,_{0.60}^{0.62}$ & $0.44\,_{0.40}^{0.47}$ \tabularnewline[.6ex]
    interaction coeff.   & $0.88\,_{0.87}^{0.90}$ & $0.99\,_{0.98}^{0.99}$ & $0.87\,_{0.86}^{0.88}$ & $0.69\,_{0.66}^{0.72}$ & $0.86\,_{0.84}^{0.87}$ & $0.99\,_{0.98}^{1.00}$ & $0.85\,_{0.84}^{0.85}$ & $0.69\,_{0.69}^{0.70}$ \tabularnewline
	\bottomrule
	\end{tabular}
\end{table}

To further quantify the visual impressions, the measures of correlation introduced in \cref{measuring correlation} are given in \cref{fit tau} for the two considered strain levels.
In this table, the average among the four combinations of microstructure/texture and load is complemented by the minimum and maximum among them.
From this table, the following observations can be made:
\begin{itemize}
    \item The visual impression of \cref{DeltaTau 110,DeltaTau 180} is reaffirmed and the highest degree of agreement is seen for interaction coefficients, followed by strengthening coefficients.
    \item Low values of the correlation coefficient $\rho$ for simplified and traditional cases indicate higher scatter (imprecision) about the best-fit line.
    \item $C_\text{b} \approx 1.0$ in the case of interaction coefficient indicate that the best-fit line agrees with the 1:1 line.
    \item Independently of the chosen parameterization, the agreement slightly decreases with increasing strain level.
    \item Similar results are obtained by all four combinations of microstructure/texture and load.
    \item MEC is always lower than $\rho_\text{c}$.
    \item $\rho_\text{c}$ indicates equal quality for the simplified and the traditional parametrizations while MEC is lower for the traditional case.
\end{itemize}

\subsection{Deformation Behavior}
\begin{figure}
	\includegraphics[width=\linewidth]{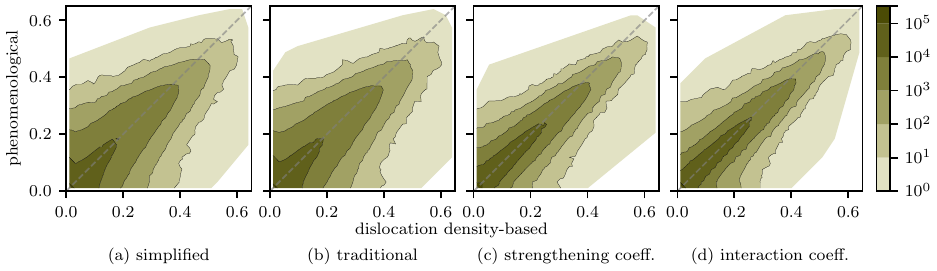}
	\caption{Density plot showing the correlation between the shear on active slip systems ($\gamma$) obtained with the different parametrizations of the phenomenological model and the dislocation density-based model at an applied strain of \qty{15}{\percent} in loading direction.}
	\label{gamma 180}
\end{figure}

For studying the impact of different parametrizations on the deformation behavior, correlation plots comparing plastic shear on active slip systems at a global strain level of \qty{15}{\percent} are presented in \cref{gamma 180}.
Overall, the correlation plots depict symmetry about the 1:1 line and the width of contour regions from this line is higher for the simplified and traditional parametrizations in comparison to the parametrizations using strengthening or interaction coefficients.

To quantify the degree of agreement, the values for the correlation metrics are reported in the \cref{fit gamma}.
An ubiquitous $C_\mathrm{b}$ value of \num{1.0} indicates that the line of best fit exactly overlaps with 1:1 line.
In this scenario, $\rho_\mathrm{c}$ is solely determined by $\rho$.
As the spread of data points about the 1:1 line increases, the correlation coefficient decreases; maximum correlation is found for the interaction coefficients.
The trend observed by $\rho$ and $\rho_\mathrm{c}$, respectively is also predicted by MEC.

\begin{table}
    \caption{Concordance correlation coefficient $\rho_{\text{c}}$ and its decomposed products: Pearson's correlation coefficient $\rho$ and bias correction factor $C_b$; and modeling efficiency coefficient MEC measuring the degree of agreement between the reference results and the different parametrizations for plastic shear $\gamma$.
    The values given in normal font are the average of all four combinations of microstructure/texture and load which are complemented by their minimum in subscript and their maximum in superscript.}
	\label{fit gamma}
	\centering
	\begin{tabular}{r c c c c c c c c}
		& \multicolumn{4}{c}{$\varepsilon \approx 0.08$} & \multicolumn{4}{c}{$\varepsilon \approx 0.15$} \tabularnewline
		\cline{2-9}
		& $\rho$ & $C_\mathrm{b}$ & $\rho_{\text{c}}$ & MEC & $\rho$ & $C_\mathrm{b}$ & $\rho_{\text{c}}$ & MEC \tabularnewline
		\toprule
		simplified           & $0.84\,_{0.81}^{0.86}$ & $1.00\,_{1.00}^{1.00}$ & $0.84\,_{0.81}^{0.86}$ & $0.68\,_{0.62}^{0.71}$ & $0.81\,_{0.77}^{0.83}$ & $1.00\,_{1.00}^{1.00}$ & $0.81\,_{0.77}^{0.83}$ & $0.61\,_{0.54}^{0.67}$ \tabularnewline[.6ex]
		traditional          & $0.84\,_{0.80}^{0.86}$ & $1.00\,_{1.00}^{1.00}$ & $0.84\,_{0.80}^{0.86}$ & $0.68\,_{0.62}^{0.72}$ & $0.80\,_{0.76}^{0.84}$ & $1.00\,_{1.00}^{1.00}$ & $0.80\,_{0.76}^{0.84}$ & $0.62\,_{0.54}^{0.68}$ \tabularnewline[.6ex]
		strengthening coeff. & $0.95\,_{0.94}^{0.95}$ & $1.00\,_{1.00}^{1.00}$ & $0.94\,_{0.94}^{0.95}$ & $0.90\,_{0.88}^{0.91}$ & $0.93\,_{0.91}^{0.93}$ & $1.00\,_{1.00}^{1.00}$ & $0.92\,_{0.91}^{0.93}$ & $0.86\,_{0.84}^{0.87}$ \tabularnewline[.6ex]
		interaction coeff.   & $0.96\,_{0.96}^{0.97}$ & $1.00\,_{1.00}^{1.00}$ & $0.96\,_{0.96}^{0.97}$ & $0.93\,_{0.92}^{0.93}$ & $0.94\,_{0.94}^{0.95}$ & $1.00\,_{1.00}^{1.00}$ & $0.94\,_{0.94}^{0.95}$ & $0.89\,_{0.87}^{0.90}$ \tabularnewline
		\bottomrule
	\end{tabular}
\end{table}

\section{Discussion}

The successful reproduction of the average stress--strain curves shown in \cref{stress-strain} by all four parametrizations clearly demonstrates that inferring the interaction parameters from macroscopic stress--strain curves is not possible.
In that context it is important to mention that this holds despite the availability of four datasets (combination of two microstructures/textures and load cases) and the simple case of an fcc lattice with a single slip family.
Hence, physics-informed parameters obtained from either experiments \cite{WuEtAl1991,BassaniEtAl1991} or small scale simulations---typically DDD---are required for an adequate parametrization.

The comparison of \cref{DeltaTau 180} and \cref{fit tau} with \cref{gamma 180} and \cref{fit gamma} shows that the use of interaction coefficients leads to a significantly higher correlation between the behavior of the phenomenological model and the reference results obtained from the dislocation density-based formulation.
This means in turn that the values of \num{1.0}/\num{1.4} typically used for the parametrization of phenomenological crystal plasticity models cannot be recommended.
The results also show that the correlation for the deformation behavior (quantified in terms of $\gamma$) between physics-based and phenomenological model is much higher for all parametrizations than the hardening behavior (quantified in terms of $\Delta \tau_\mathrm{crit}$).
Related to that, it is also observed that the improvement in correlation is much higher for the hardening behavior than for the deformation behavior.
The reason for that can be explained by the fact that in monotonous loading only a subset of slip systems is active, c.f. \cref{fraction active}.
In the limiting case of having only one active slip system, all parametrizations are equivalent due to $h_0 = 1.0$, see \cref{h values}.
The same holds for $h_\mathrm{copla}$, where only the simplified parametrization uses a value of \num{1.4} instead of \num{1.0}.
Hence, despite a poor agreement for hardening---which concerns also inactive slip systems---the predicted deformation behavior is in rather good agreement.
The situation would be very different for load path changes, e.g. when simulating cycling loading, and texture evolution due to large deformations \cite{SedighianiEtAl2021} as that would typically lead to the activation of previously inactive systems.
In this case a significant difference in the deformation behavior is expected between the phenomenological model with the three ``wrong'' parametrizations and dislocation density-based models.
It should be noted that the models employed in this study do not include a back stress term and are therefore not particularly suited for load path changes that may result in slip reversals.
However, it is anticipated that the results from this study are transferable to phenomenological formulations designed for load path changes \cite{WollmershauserEtAl2012}.

The use of interaction coefficients has also consequences for the feasible ranges of other parameters of the models.
In particular, it is obvious that $h_\mathrm{coli} = 5.51$ leads to pronounced hardening of slip systems that are collinear to active systems.
That mandates sufficiently high saturation value $\tau_\mathrm{crit}^\infty$ to avoid coming close to the saturation value at low to moderate strain levels.
It should be noted that this issue is partly caused by the use of constant values for the interaction parameters in \cref{dislocation}, which is a simplification that does not take into account interaction strength decreases with increasing dislocation density \cite{DevincreEtAl2006,QueyreauEtAl2009}.

%%%%%%%%%
% Conclusion
%%%%%%%%%

\section{Conclusion and Outlook}
It is shown that use of interaction coefficients determined by means of DDD simulations improves the parametrization of phenomenological crystal plasticity models.
More specifically, using the values obtained by DDD simulations increases the agreement between reference results obtained with a dislocation density-based crystal plasticity model in comparison to typically used values (``1.0/1.4'') for phenomenological simulations.
The investigations have been performed on the example of Nickel, an fcc material.
Nevertheless, it is anticipated that the findings can be transferred to materials with other crystal structures.
Both, hardening, quantified in terms of the increase of CRSS on all slip systems, and deformation behavior, quantified in terms of the plastic shear on active slip systems, have been compared.
The differences are significantly larger for the hardening behavior than for the deformation behavior
Hence, the use of physics-informed hardening parameters is particularly important for simulations involving load path changes, e.g. in the case of cyclic loading, that lead to the activation of previously inactive (but hardened) slip systems.

Since DDD parametrizations exists for most fcc \cite{MadecEtAl2003,DevincreEtAl2006,KubinEtAl2008,MadecEtAl2017,AkhondzadehEtAl2020} and body-centered cubic \cite{QueyreauEtAl2009,MadecEtAl2017} metals and for some hexagonal close packed \cite{Devincre2013,BertinEtAl2014} materials, this findings allow for easy improvements of phenomenological crystal plasticity model parametrizations.
It is also hypothezized that the findings presented here are also applicable to other phenomenological hardening models, e.g. those suitable for load path changes \cite{WollmershauserEtAl2012}.

To further improve the capabilities of the phenomenological formulation, it might be required to decrease the magnitude of the interaction parameters with increasing deformation.
However, implementing the approach presented by \cite{DevincreEtAl2006} is not straight forward as it requires to have the dislocation density as an internal variable.
Furthermore, it must be questioned whether further complications to the phenomenological model, which is preferentially used because of its simplicity, are desired.

%\section*{Stray Text Blocks}
% Voce Kocks-Mecking equivalence:\cite{EstrinEtAl1984}\cite{PatilEtAl2021}

% Note that this is a simplification, e.g. stress dependence \cite{QueyreauEtAl2009}. ToDo: check if correct citation.

% Related publications, probably not relevant enough for citation
% \cite{MadecEtAl2002} only isotropic (matches experiments)
% \cite{MadecEtAl2002a} only isotropic (matches experiments)
% \cite{MadecEtAl2002b} only isotropic (matches experiments)
% \cite{MadecEtAl2004} interaction matrix bcc, newer work is available
% \cite{MadecEtAl2008} ternary conﬁguration containing a second-order junction (bcc/fcc)

% \cite{MonnetEtAl2004} no concrete values (more qualitative)

% \cite{DevincreEtAl2008} mean free path, cites DevincreEtAl2006

\section*{Acknowledgements}

This research was financially supported by Internal Funds KU Leuven.

\section*{Conflict of Interest}

The authors declare no conflict of interest.

\section*{Author Contributions}

\textbf{Nikhil Prabhu} and \textbf{Martin Diehl}: Conceptualization; Methodology; Software; Validation; Formal analysis; Investigation; Data Curation; Writing – original draft; Writing – review and editing; Visualization.
\textbf{Martin Diehl}: Resources; Supervision; Project administration; Funding acquisition.

\section*{Data Availability Statement}

The data to reproduce the simulations with DAMASK \cite{RotersEtAl2019,OttodeMentockEtAl2025} conducted for this study is available on Zenodo \cite{prabhu_2025_14799861}.

\AtNextBibliography{\small}
\printbibliography

% Figures/tables and captions
% Permission statements are required for all figures reproduced or adapted from previously published articles/sources. Please also ensure that all necessary permissions to reproduce images have been received
% Please remove these statements for original figures
%\begin{figure}
%  \includegraphics[width=\linewidth]{placeholder-image.png}
%  \caption{Figure 1 caption goes here. Reproduced with permission.\textsuperscript{[Ref.]} Copyright Year, Publisher. }
%  \label{fig:boat1}
%\end{figure}

%\begin{table}
% \caption{Table 1 caption}
%  \begin{tabular}[htbp]{@{}lll@{}}
%    \hline
%    Description 1 & Description 2 & Description 3 \\
%    \hline
%    Row 1, Col 1  & Row 1, Col 2  & Row 1, Col 3  \\
%    Row 2, Col 1  & Row 2, Col 2  & Row 2, Col 3  \\
%    \hline
%  \end{tabular}
%\end{table}

\end{document}